\documentclass[12pt]{iopart}

\usepackage{graphicx}

\begin{document}

\title{Graphene as a non-magnetic spin-current lens}

\author{F S M Guimar\~aes$^1$, A T Costa$^1$, R B Muniz$^1$ and M S Ferreira$^2$} 
\address{$^1$ Instituto de F\'{\i}sica, Universidade Federal Fluminense, Niter\'oi, Brazil}
\address{$^2$ School of Physics, Trinity College Dublin, Dublin 2, Ireland}
\ead{ferreirm@tcd.ie}

\begin{abstract}
In spintronics, the ability to transport magnetic information often depends on the existence of a spin current traveling between two different magnetic objects acting as source and probe. A large fraction of this information never reaches the probe and is lost because the spin current tends to travel omni-directionally. We propose that a curved boundary between a gated and a non-gated region within graphene acts as an ideal lens for spin currents despite being entirely of non-magnetic nature. We show as a proof of concept that such lenses can be utilized to redirect the spin current that travels away from a source onto a focus region where a magnetic probe is located, saving a considerable fraction of the magnetic information that would be otherwise lost.
\end{abstract}


\maketitle

The field of spintronics is based on ingenious ways of controlling not only how the electron charge flows across nanoscaled materials, but also how the electron spin propagates within the same environment \cite{spintronics-review}. One of the key challenges in the field is to find materials with large spin-coherence lengths, so that the information contained in the electron spin is not lost after a short propagation distance. Carbon-based materials such as nanotubes and graphene have been in the scientific limelight for over a decade now and among their many interesting properties is the fact that they have unusually long spin-coherence lengths \cite{slength1,slength2,slength3,slength4,lundeberg}. Therefore, it seems like a perfect match to study spintronics in such carbon-based materials, as confirmed by the growing number of studies on the subject \cite{esquinazi, jpcm04, prb04, coupling1, cohen, McEuen, kim, ah-review, suppression, carbon}.

One of the peculiarities of graphene is the linear dispersion relation for electrons at the Fermi level region, which makes electronic wave packets travel dispersionlessly. This is contrary to what is commonly expected of typically dispersive quantum particles described by packets with wave-vector components that travel at different speeds. Such a lack of dispersion in graphene suggests that its electrons must evolve very much like typical classical waves. As a result, analogies with optical systems have been made in an attempt to widen the applicability of graphene \cite{falko,lenses-prl}. Here we propose that this optical analogy be further explored to generate graphene-based lenses that are entirely non-magnetic and yet capable of concentrating spin currents onto a focal point.  

Spin currents correspond to a net flow of electron spins that may or may not coexist with charge current.  One way of producing such currents is by pumping spins into a non-magnetic metal through the precession of an adjacent magnetization \cite{Tser, bauer-e-cia}. In this case, angular momentum from the moving magnetization is transferred to the conduction electrons, creating a spin disturbance that propagates throughout the conducting material. It is noteworthy that spin currents may also be generated by a rotating magnetic field, as reported in reference \cite{wangchan}, with no need of magnetic atoms being attached to the graphene lattice structure. Let us emphasize that this spin flow is produced without an applied voltage and involves no net electrical current. Moreover, the traveling spin current can in principle be used to remotely excite other magnetic units located a long distance from the source of magnetic precession, so that the conducting material acts as a waveguide of magnetic information\cite{waveguide, sct}. The obvious drawback of using this method for transporting information is that because the initially localized precession tends to travel omni-directionally, a good fraction of this information is lost if it is not reabsorbed by another magnetic moment. Ideally, one would like to redirect the spin current and focus it towards a magnetic object capable of absorbing the spin-precession energy that would be otherwise lost. 

The ease with which graphene is gated suggests that it is straightforward to make its dispersion relation spatially dependent according to the gate shape. Because the gate strength defines the velocity with which electrons travel across a material, one can envisage a way in which spatially dependent refraction indices are engineered for electrons propagating in graphene. Both refraction and diffraction phenomena affect the propagation direction of a wave and are able to induce focussing patterns. In this work, we show with a fully quantum mechanical approach how spin-current focussing may be achieved by dividing a large graphene region into two parts, one of which is under the action of a gate voltage. It is worth stressing the enormous benefit of controllably redirecting the spin-current in order to minimize the loss of magnetic information in spintronic devices. 

\begin{figure}
\begin{center}
\includegraphics[width = 8.cm] {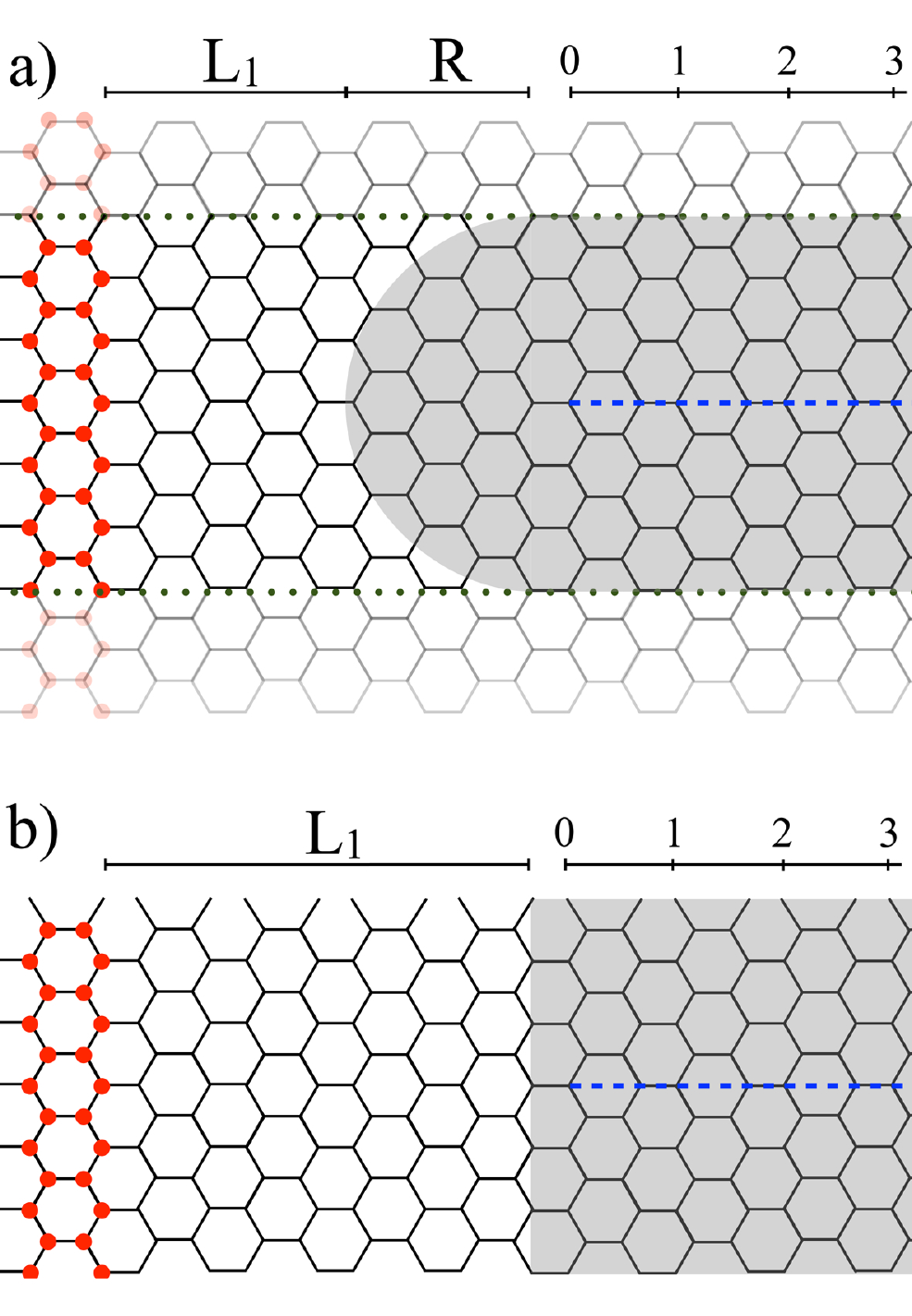}
\end{center}
\caption{Schematic diagrams representing the structures under study. (a) An infinitely large graphene sheet is under the effect of a gate voltage that acts only in a limited region of space (represented by the gray area). The gated region has a boundary with a curvature defined by the radius $R$ that lies at a distance $L_1$ from the the magnetic atoms. The section delimited by the two horizontal dotted lines is the unit cell used in representing the system with lateral periodic boundary conditions. (b) Flat boundaries are easily accounted for in the limit $R \rightarrow \infty$. Dashed lines represent the locations where the spin disturbance will be probed. }
\label{figure1}
\end{figure}

Let us consider a graphene sheet with a few magnetic atoms substitutionally inserted into its hexagonal atomic lattice. Magnetic atoms, usually transition metal elements, are known to display sizable magnetic moments \cite{arkady-prl, vene-prb, portal-NJP} when inserted in graphene and will be used here as the source of spin current. Other objects such as magnetic nanoparticles, ferromagnetic substrates or even vacancies \cite{yazyev} can also be used but substitutional atoms are by far the simplest and yet capture the essence of the phenomenon we intend to describe. We also consider one stripe of magnetic atoms as schematically illustrated in figure \ref{figure1}. While a pristine stripe of substitutional impurities is unlikely to exist, here it serves the purpose of (a) testing how the system responds to the size of the magnetic material responsible for generating the spin current and (b) inducing spin currents that quickly develop into plane waves. Furthermore, we assume the system to have a curved boundary separating two different media, one of which under the action of an external gate voltage $V_g$ which effectively controls the electron density in the region. Figure 1(a) shows a schematic diagram of the system under consideration. Note that the boundary between the gated (gray) and non-gated (white) regions is curved and chosen to be circular with a radius $R$. The obvious problem of considering such a setup is the difficulty of describing this system without paying incredibly hefty computational prices given the enormous number of atomic sites required. For that reason we choose to work with finite-width sections of graphene with lateral periodic boundary conditions (PBC), where the repeated unit cell is the part of figure \ref{figure1}(a) delimited by the two horizontal dotted lines. Note that flat boundaries are also straightforwardly considered by taking the limit $R \rightarrow \infty$, as shown in figure \ref{figure1}(b). 

As previously mentioned, the pumped spin current is generated by the precession of a magnetization in contact with a conducting medium. We assume that the magnetization is originally in equilibrium, pointing along an arbitrary z-direction, and that it is set into precession by a time-dependent transverse field $h_\perp(t)$. In practical terms, there are different ways of producing as well as probing this time-dependent perturbation \cite{bauer03,wees-nature,wees}. To determine how graphene transports spin currents, we must assess how such a localized magnetic excitation propagates across the structure. When such a spin current flows across the normal metal it creates a time dependent local spin disturbance that may be described by the transverse spin susceptibility. If the amplitude of $h_\perp(t)$ is sufficiently small, one may use linear response theory to relate the spin disturbance at site $m$ with the driving field applied at site $\ell$. The relation for a harmonic perturbation with amplitude $h_0$ is given by $\delta\langle S^+_m(t) \rangle = g\mu_Bh_0 e^{-i\omega t}\chi^{+-}_{m,\ell}(\omega)/2$, where $S^+_m = S^x_m+iS^y_m$, and $S^{x,y}_m$ represent the local transverse spin components at site $m$. It follows that $|\delta\langle S^+_m(t) \rangle| = g\mu_Bh_0 |\chi^{+-}_{m,\ell}(\omega)|/2$ and, therefore, the off-diagonal transverse spin susceptibility provides the amplitude of the spin disturbance caused by the spin current that flows through the normal metal. Here, we investigate how this excitation disturbs the spin balance of the system not only where the magnetic atoms are located but also, and more importantly, how the local spin dynamics is affected within graphene.

To calculate the spin susceptibility \cite{bmcm, muniz-mills} one needs the Hamiltonian describing the electronic structure of the unperturbed system, which we assume is given by  $\hat{H} =   \sum_{i,j,\sigma} \gamma_{ij} \, \ {\hat
c}_{i\sigma}^\dag \, {\hat c}_{j\sigma} + \sum_{\ell,\sigma}  {U \over 2}  \, {\hat n}_{\ell \sigma} \, {\hat n}_{\ell {\bar \sigma}} + \hat{H}_Z$. Here, $\gamma_{ij} $ represents the electron hopping between nearest neighbor sites $i$ and $j$,  $\hat{c}_{i\sigma}^{\dag}$ creates an electron with spin $\sigma$ at site $i$, the sum in $\ell$ is over the sites occupied by magnetic atoms, $\hat{n}_{\ell \sigma} = \hat{c}_{\ell \sigma}^{\dag} \hat{c}_{\ell \sigma}$ is the corresponding electronic occupation number operator, and $U$ represents an effective on-site interaction between electrons in the magnetic sites, which is neglected elsewhere. With the lateral PBC imposed, the Hamiltonian matrix is considerably reduced. Finally, ${\hat H}_Z$ plays the role of a local Zeeman interaction that defines the $\hat{z}$-axis as the equilibrium direction of the magnetization. The Hamiltonian parameters can be obtained from density-functional-theory calculations so that the electronic structure of the doped system is well described \cite{charlier-prl04, rocha}. Although the presented results are for Mn atoms, other substitutional magnetic impurities may be employed. In our calculations we fix the Fermi energy $E_{F}=0$, and use $\gamma_{i,j} = 2.7$ eV for the nearest-neighbour hopping terms. We take the number of $d$-electrons in the Mn sites $n_m = 1$, $U=20$ eV, and assume a local Zeeman energy splitting of $1$ meV, which is of the order of magnitude of magnetic anisotropy energies found for transition metal atoms adsorbed to metallic surfaces \cite{khaje,hongwu}. For simplicity we consider the electronic hopping between carbon atoms to be the same as the hopping between carbon and the magnetic impurity. Within a five-fold degenerate orbital scheme, this requires a large value of U to obtain a reasonable value for the Mn magnetic moment. Spin-orbit coupling, which has been shown to play a role in the case of magnetic adsorbates\cite{guinea}, is neglected in the substitutional impurity case for being very small compared to the other relevant energy scales. 

\begin{figure}
\begin{center}
\includegraphics[width = 8.cm] {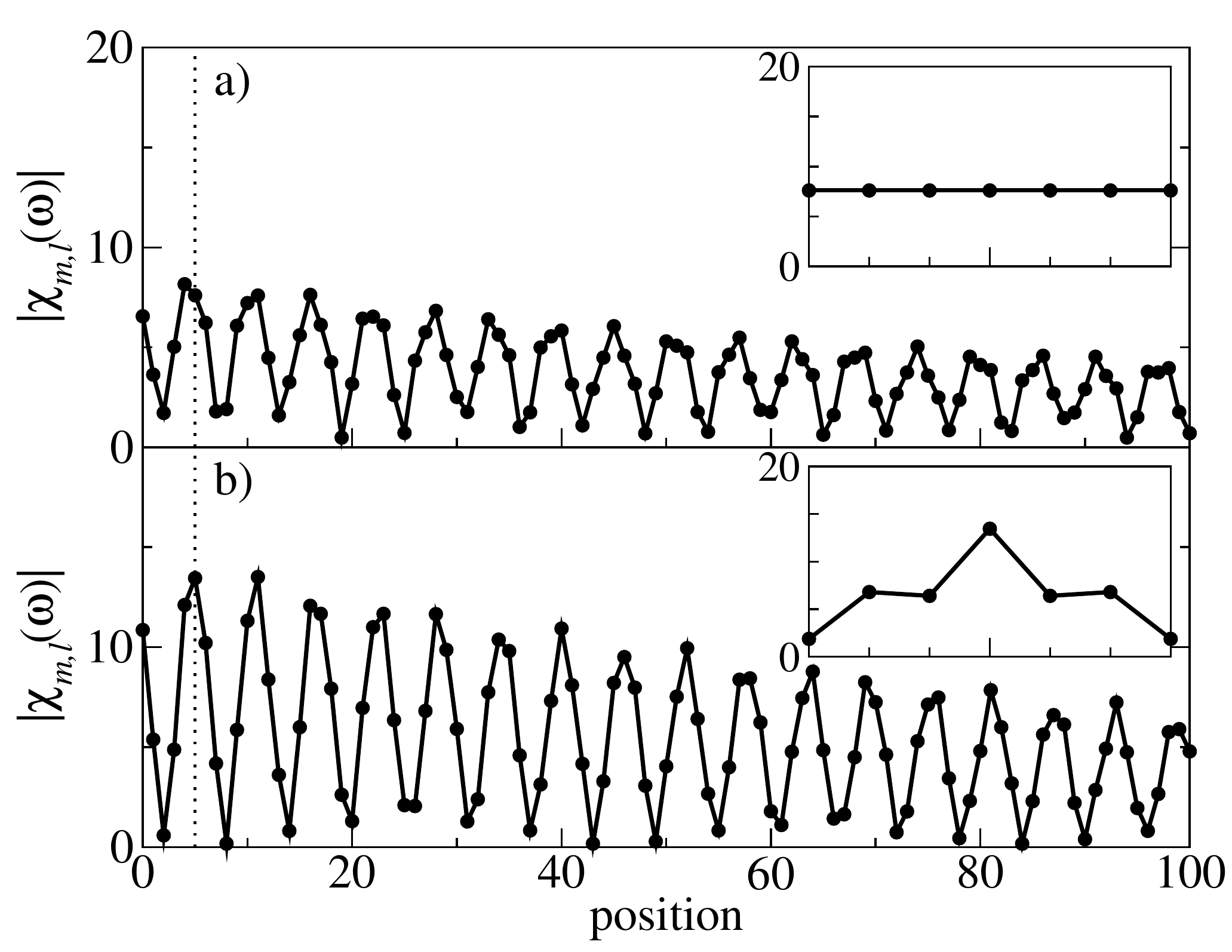}
\end{center}
\caption{Magnitude of the spin susceptibility calculated at resonance frequency for different positions along the horizontal dashed lines shown in figure \ref{figure1}. The insets depict the spin susceptibility probed at the carbon sites situated along a vertical line located at position $5$ of the horizontal axis shown in figure \ref{figure1}. The dotted lines in figures (a) and (b) mark the position of these vertical lines. All graphs are for $V_g = 2.4 \, {\rm eV}$. (a) Results for a flat boundary ($R \rightarrow \infty$). (b) Results for a curved boundary ($R=3 a$)}
\label{figure2}
\end{figure}

The time-dependent transverse spin susceptibility is defined as $\chi_{m,\ell}(t) = -{i \over \hbar} \Theta(t)\langle[{\hat S}_m^+(t),{\hat S}_\ell^-(0)]\rangle$, where $\Theta(x)$ is the heaviside step function, and ${\hat S}_m^+$ and ${\hat S}_m^-$ are the spin raising and lowering operators at site $m$, respectively. The indices $\ell$ and $m$ refer to the locations where the field is applied and where the response is measured, respectively. In our case, we induce a precession of the magnetic moments on sites $\ell$ and we wish to probe the resulting spin disturbance at an arbitrary site $m$. Note that the probing site may or may not contain another magnetic object without affecting the conclusions here obtained. The effect of a magnetic object occupying the probing position simply amplifies the resulting spin disturbance. In what follows we present results for which there are no magnetic sites on the probing region. In this case, this precession response is fully described by $\sum_\ell \chi_{m,\ell}(t)$. Within the random phase approximation, this susceptibility may be expressed in frequency domain, and in matrix form it is given by $\chi(\omega) = [1 + \chi^0(\omega)\, U]^{-1} \, \chi^0(\omega)$, where $\chi^0$ is the Hartree-Fock susceptibility given by
\begin{eqnarray}\nonumber
\chi_{m,\ell}^{0}(\omega) = \frac{i}{2\pi} \int_{-\infty}^{\infty}d\omega^{\prime}f(\omega^{\prime})&\left\{[ g_{\ell m}^{\uparrow}(\omega^{\prime}) -  g_{\ell m}^{- \uparrow}(\omega^{\prime})] g_{m\ell}^{\downarrow}(\omega^{\prime} + \omega)\right.\\
&\left.+ [ g_{m\ell}^{\downarrow}(\omega^{\prime}) -  g_{m\ell}^{- \downarrow}(\omega^{\prime})] g_{\ell m}^{- \uparrow}(\omega^{\prime} - \omega)\right\}\ .
\end{eqnarray}
Here, $g_{m\ell}^{\sigma}(\omega)$ and $g_{m\ell}^{- \sigma}(\omega)$ represent the time Fourier transforms of the retarded and advanced single-particle propagators for an electron with spin $\sigma$ between sites $m$ and $\ell$, respectively, and $f(\omega)$ is the Fermi-Dirac distribution function. Following reference \cite{muniz-mills}, it is possible to extend some those integrals to the complex plane, thereby avoiding a great deal of numerical difficulties. The one electron propagators are obtained by a well-established numerical methods \cite{sancho,mathon}. $\chi^0$ describes single-particle (Stoner) excitations whereas the enhanced susceptibility $\chi$, calculated within the RPA, takes into account correlated electron-hole pairs, that correspond, in an extended ferromagnet, to collective modes known as spin waves \cite{vignale}.

We have recently studied the spin disturbance in carbon nanotubes as a function of time and have identified that pulsed magnetic excitations travel across these materials with their corresponding Fermi velocities, with very little deformation and with adjustable levels of attenuation, something that has been attributed to the distinctive linear dispersion relation displayed by nanotubes \cite{waveguide}. Graphene is no different in this regard and, for having similar linearities in its dispersion relation, should behave in exactly the same way. In fact, we have demonstrated this similarity with graphene ribbons by showing that localized spin excitations also propagate dispersionlessly and with negligible loss. Furthermore, we have also argued that a gated region between the magnetic object generating the spin current and the locations where the spin disturbance is probed may be capable of modulating the spin precession between on- and off-states, giving rise to a transistor behavior for the pumped spin current that functions without the need of a bias to drive the current\cite{sct}. 

Rather than studying the time-resolved spin excitations\cite{waveguide}, in this paper we focus on the frequency-dependent response. The key quantity for us is therefore $\vert \chi_{m,\ell}(\omega)\vert$ because it is proportional to the amplitude of the spin disturbance at site $m$ due to a time-dependent transverse magnetic field applied at site $\ell$. In other words, it corresponds to the precession amplitude acquired by the electron spin localized at an arbitrary site $m$ due to a harmonic perturbation of frequency $\omega$ produced at the magnetic site $\ell$. Unsurprisingly, when investigated as a function of frequency this quantity displays a very distinctive peak close to its Larmor-frequency resonance. Without any loss of generality, we use this value as our choice of excitation frequency.

Figure 2(a) shows the results for $\vert \chi_{m,\ell}(\omega)\vert$ probed at different sites $m$, all of which following the horizontal dashed line depicted in Figure 1(b). This result corresponds to a flat interface ($R=\infty$) separating the two different media, one of which under the action of a gate voltage $V_g = 2.4 \, \rm{eV}$. There is a noticeable oscillation in the amplitude of the spin precession as a function of the probing position along the axis of the ribbon, which can be ascribed to the inevitably oscillatory spin polarization induced by the presence of magnetic objects in contact to conduction electrons. As previously mentioned, because the precession amplitude is comparatively increased for larger spin moments the same oscillation appears in the susceptibility results. The inset of figure \ref{figure2}(a) shows that the response is fairly homogeneous when probed along a direction perpendicular to the ribbon axis, located at position $5$ of the horizontal axis shown in figure \ref{figure1}(b). This means that the spin precession caused by the spin current travels into the gated medium as a plane wave without favoring any subpart of the system. 

\begin{figure}
\begin{center}
\includegraphics[width = 10.cm] {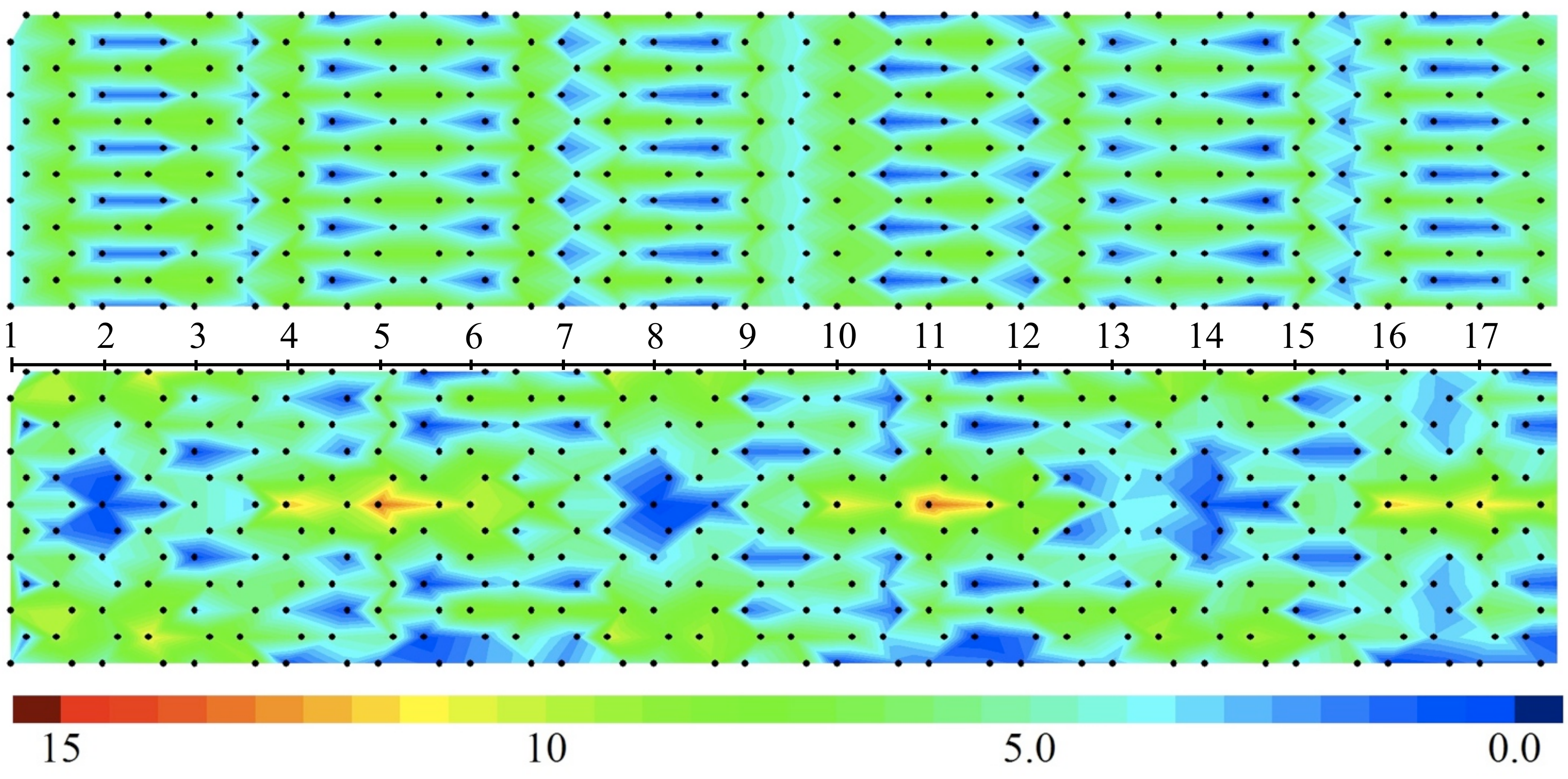}
\end{center}
\caption{Color-coded plot of the spin susceptibility as a function of the position within the gated region depicted by the gray areas in figure \ref{figure1}. The Top (Bottom) panel corresponds to a flat (curved) boundary with parameters specified at the captions of figure \ref{figure2}. }
\label{figure3}
\end{figure}

A different response in the precession amplitude is found in figure \ref{figure2}(b) when the flat interface is replaced with a curved surface of radius $R = 3 a$, where $a$ is the lattice parameter of graphene\cite{obs2}. The gate voltage is the same as the one used in figure \ref{figure2}(a) but the amplitude of the oscillations is nearly twice as large. The oscillatory spin polarization induced by the proximity to a magnetic object is known to be primarily dependent on the details of the medium in which the polarization occurs, indicating that a change in the interface shape cannot enhance the induced moments very significantly. Therefore, the increase in the amplitude of $\vert \chi_{m,\ell}(\omega)\vert$ is likely to result from a focussing process. A more convincing evidence that the spin current is being focussed by the curved interface is shown in the inset of figure \ref{figure2}(b), which displays $\vert \chi_{m,\ell}(\omega)\vert$ probed on a line that is perpendicular to the ribbon axis, located at position $5$ of the horizontal axis shown in figure \ref{figure1}(a). A maximum is clearly seen at the central region of the ribbon, indicating that the peaks seen as a function of the axial positions are not distributed homogeneously across the ribbon width but are concentrated at a small region of space acting as a focus for the spin current. This is even more conclusively shown in figure \ref{figure3}, where 2-dimensonal color plots representing the precession amplitude as a function of the probing position in the graphene ribbon is displayed. The upper panel corresponds to the flat interface and is rather featureless. That confirms our earlier statement that the precession amplitude is rather homogeneous for flat surfaces. For the case of curved interfaces, however, the bottom panel of figure \ref{figure3} shows a region where the spin susceptibility reaches a localized maximum acting as a center where the spin current is focussed. It is worth noting that this region is surrounded by other areas where the spin susceptibility is very low. This is understood when we compare the bottom and upper panels of figure \ref{figure3}, indicating that the precession energy that would be homogeneously spread across the ribbon width in the flat-surface case is redirected towards a focus when in the presence of a curved interface. As a result, a large part of the magnetic information contained in the precession motion of the electron spins would not be probed and would be completely lost. By playing with the combination of gate voltage and geometry of the interface separating two parts of a graphene sheet, we can minimize this information loss. Finally, it is worth stressing that the results reported here are very robust and remain qualitatively the same for a wide range of diameters and voltages. This opens a number of possibilities for using the energy contained in the precession of magnetic moment as a source of magnetic information. The fact that lenses can be used to concentrate the spin current onto a small region of space is ideal to amplify spin signals that could be otherwise very weak. 

In summary, we propose that the spin current originated from a precessing magnetization in contact to a graphene sheet can be controllably redirected towards a focus region, which could reduce the loss of magnetic information stored in the spin dynamics of such systems.  Furthermore, by probing the spin current at the focus region one is likely to enhance the sensitivity of any spin pumping device of this type.


\begin{thebibliography}{36}
\bibitem{obs2} The small value of $R$ is not a limitation and the results here reported are unchanged as the structure is scaled up.
\end{thebibliography}
\end{document}